\documentstyle[12pt]{article}          
\catcode`\@=11
\@addtoreset{equation}{section}

	\newdimen\eqskip
	\newdimen\txtskip
	\baselineskip=\txtskip
\def	\be		{\begin{equation}}
\def	\ee		{\end{equation}}
\def	\ba		{\begin{eqnarray}}
\def	\ea		{\end{eqnarray}}

\def	\=		{\;=\;}
\def	\frac		#1#2{{#1 \over #2}}

\def	\to		{\rightarrow }

\def	\as		{\mbox{$\alpha_s$}}

\def	\jpsi		{\mbox{$J/\psi$}}
\def	\psitwos	{\mbox{$\psi(2S)$}}

\def	\ppbar		{\mbox{$\bar p p$}}
\def    \pt	        {\mbox{$p_t$}}

\def    \ptqq	        {\mbox{$p_t^{Q\bar Q}$}}

\def    \et	        {\mbox{$E_t$}}
\def \dphi{\mbox{$\Delta\phi$}}

\def \lqcd  {\mbox{$\Lambda_{QCD}$}}

\def \lff   {\mbox{$\Lambda^{2-loop}_{4}$}}

\begin{document}
\nopagebreak
{}.
\begin{flushright}
IFUP-TH 60/93 \\
November 1993
\end{flushright}
\begin{center}
{\Large { \bf \sc Recent Progress in the Theory of Heavy Flavor Production
\footnote{Invited talk presented at the 9th Topical Workshop on
Proton-Antiproton Collider Physics, 18-22 October 1993, Tsukuba, Japan.}}}
\vskip .3cm
{Michelangelo L. Mangano}
\\
{\it INFN, Scuola Normale Superiore and \\
    Dipartimento di Fisica, Pisa, Italy}\\
\end{center}
\nopagebreak
\begin{abstract}
{We review some recent results on heavy quark production in high energy
hadronic collisions. We will discuss in particular the status of production
cross sections for bottom quarks and charmonium states and will present some
studies on the production of bottom and charm jets, at the inclusive level and
in association with electroweak gauge bosons.}
\end{abstract}
%
%
\section{Introduction}
Heavy quark production in high energy hadronic collisions consitutes a
benchmark process for the study of perturbative QCD \cite{nason}.
The comparison of
experimental data with the predictions of QCD provides a necessary check that
the ingredients entering the evaluation of hadronic processes (partonic
distribution functions and higher order corrections) are under control and can
be used to evaluate the rates for more exotic phenomena or to extrapolate the
calculations to even higher energies. The estimates of production rates for the
elusive {\em top} quark rely on the understanding of heavy quark
production properties within QCD.

Much progress has been made in the field since the last \ppbar\ Collider
Workshop in 1989, both experimentally and theoretically. Being impossible to
fit in the time allowed a complete review of these developments, I will shortly
summarise them here before proceeding with the recent material,
for the benefit of the unaware reader.

At the time of the last Workshop, one of the outstanding issues in the field
was the disagreement between the  measurement of the CERN experiment UA1 of the
B cross section at large \pt\ \cite{ua1_b1} and the first theoretical
calculation of this distribution at the next-to-leading order in QCD (a.k.a.
NDE, \cite{nde}). No results from the Collider Detector at Fermilab (CDF) were
available as yet. Since then, new data from UA1 have come into nice agreement
with the theoretical QCD expectations \cite{ua1_b2}, and an independent QCD
calculation  has been performed by a different group \cite{vanNeerven},
providing an important confirmation of the NDE results.

The dust on the UA1 new analysis had barely settled, when the success of QCD in
describing the B production cross section was put into serious embarassement by
the first results from CDF \cite{cdf_b1,cdf_psi}, indicating a significant
discrepancy between the data and the same theoretical prediction working so
well at the lower CERN energies.

This disagreement at higher energy, whose possibility had already been
anticipated in the original NDE article, was justified qualitatively
by the inadequacy of a fixed order calculation in perturbation
theory (PT) applied  to processes probing
initial state gluons at very small Bjorken-$x$. In this regime, in fact, large
corrections of order $[\as \log(1/x)]^n \sim [\as \log(s/m^2)]^n$  appear at
higher orders in PT. Some of the most interesting
theoretical papers in this period have been devoted to the development of
techniques designed to resum these corrections \cite{smallx}. While it became
clear that indeed these effects can increase the total B cross section by
significant factors at asymptotic energies, the increase from 0.63 to 1.8 TeV
between the CERN and FNAL energies was seen to be insufficient to explain the
factor of 3 or more observed rate discrepancy.

Alternative explanations have been attempted, most notably the possibility
\cite{berger}\ that the gluon distribution function as determined from standard
fits of low energy data is not known sufficiently well to allow a reliable
extrapolation to the small values of $x$ involved in B production at the
Tevatron.
While there is here some room for improvement, as shown in \cite{berger}, it is
my view \cite{mlmbpsi}\ that with the most recent data and analyses of
structure functions \cite{newmrs,cteq}\ it would be hard to account for a
change in the expected B cross section as drastic as required by the first CDF
data.

In the attempt to further probe the dynamics of heavy quark production, a full
NLO calculation of the $Q\bar Q$ correlations has also been carried out in the
recent past \cite{mnr}. Quantities predicted by this calculation  (a.k.a. MNR)
are, among others, the \pt\ distribution of the quark pair system (\ptqq) and
the shape of the \dphi\ distribution, \dphi\ being the difference in azimuth
between the quarks in the plane transverse to the beam axis. Measurements of
this distribution have been carried out both by UA1 \cite{geiser}\ and by CDF
\cite{cdf_dphi}, indicating good agreement with the expectations of NLO QCD.
In this area, a recent study has improved the theoretical calculation of the
mass and \ptqq\ correlations by including the effects of multiple soft gluon
emission from the initial state \cite{berger2}.

The corrections to the inclusive \pt\ distribution due to multiple soft gluon
emission have also been calculated, in an important recent paper by Laenen et
al. \cite{laenen}. Since these effects contribute terms of order $[\as
\log(m/\ptqq)]^n$, they turn out not to affect significantly the bottom cross
section, while they can be of importance in the case of the more massive top
quark, which is produced much closer to threshold and for which soft gluon
exchanges are more important.
A complete discussion of the results for the top can be found in
ref.\cite{laenen2}.

Parallel to the theoretical developments, signficant improvements have taken
place on the experimental side. We can now benefit from the presence of the new
detector, D0, whose preliminary B results have been shown also at this meeting
\cite{D0}. Furthermore, the upgrade of the CDF detector with a silicon
microvertex detector, capable of reconstructing the secondary vertices
from B decays, is increasing CDF's ability to identify B's with reduced
backgrounds \cite{CDFKEK}. This is fundamental, for example, in separating the
component of the \jpsi's produced directly from those produced in the decay of
B states.

The new preliminary results on the \pt\ distribution of fully reconstructed B
mesons, presented here by D. Crane \cite{CDFKEK}, suggest a decrease of the
measured cross section in the small \pt\ region when compared to what
previosuly observed \cite{cdf_b1}.  I collect in Fig.~\ref{fbcdf}\ the
available CDF data and compare them with the NLO QCD calculation
\cite{nde,vanNeerven}.
\begin{figure}
\begin{minipage}[t]{8.5cm}
\vskip 6.5cm
\caption[]{\tenrm \baselineskip=12pt
Integrated $b$ quark \pt\ distribution at 1.8 TeV. CDF data versus NLO QCD.
\label{fbcdf}}
\end{minipage}
\hskip 0.5cm
\begin{minipage}[t]{8.5cm}
\vskip 6.5cm
\caption[]{\tenrm \baselineskip=12pt
Diagrams for the production of a \jpsi: direct (top) and
via gluon fragmentation (bottom).
\label{ffragpsi}}
\end{minipage}
\end{figure}
I pushed to the limit the input parameters of the
theoretical calculation, namely the renormalization/factorization scale (set to
$\mu=\mu_0/4$, with $\mu_0=\sqrt{m^2+\pt^2}$), and the value of \lqcd\ (set to
\lff=275 MeV for the MRSD0 PDF set \cite{newmrs}). This choice of parameters
still provides an acceptable upper limit to the theoretical expectation. As
clear from the plot, if the new data were confirmed in the final analyses by
CDF and D0, the discrepancy with theory would have disappeared, though
at the expense of selecting extreme values for the theoretical input
parameters.

The situation with the B cross section at the Tevatron is therefore what I
would define as {\it fluid}: new data from HERA in the small $x$ and large
$Q^2$ region will soon provide more reliable parametrisations of the gluon
densities, and new data and analysis techniques from CDF and D0 will further
reduce the experimental uncertainties of the measurements.

In the rest of this report I will concentrate on two items:
\begin{itemize}
\item recent progress in the calculation of charmonium production
\item studies of charm and bottom jet production, both at the inclusive level
and in association with electroweak gauge bosons.
\end{itemize}
The interest in these two subjects arises from different observations.
The driving force for the former is directly related to the measurement of the
B cross sections: B decays are in fact an important component of the observed
rates of charmonium states, \jpsi, $\chi$ and \psitwos. In the case of the
\psitwos, for example, it is currently believed from theoretical calculations
\cite{gms,mlmbpsi}\ that almost 100\% of the observed events result from B
decays. Large fractions are also expected for
the \jpsi. Independent measurements of the B cross section, and knowledge of
the decay branching ratios, will therefore provide by subtraction
a measurement of the direct sources of charmonium, which can then be tested
against our QCD production models.

Interest in the latter subject comes from several sources. To start with, the
study of heavy quark jets (namely jets originating from the evolution of a
promptly produced $Q$ or jets containing a heavy quark produced during the
gluon shower evolution) provides yet another test of our understanding of QCD.
Furthermore, b-jets produced in association with $W$ bosons provide powerful
signatures for the detection of the top quark \cite{KEKTOP}. It is therefore
important to understand all possible sources of backgrounds induced by direct
production of heavy quark jets. The use of different data samples which are
known not to be contaminated by a possible top, but which do contain at some
level bottom or charm jets, allows to explore the performance of the heavy
quark tagging techniques, and to determine the reliability of the theoretical
expectations.
Finally, associated production of EWK bosons and heavy quarks has been
suggested as a possible tool to study the heavy flavour content of the proton
\cite{halzen}.

A more thorough discussion of all of these questions will hopefully
appear soon \cite{mlm94}.

Before concluding this Introduction, let me just mention an additional couple
of very recent theoretical developments in the theory of heavy quark production
in hadronic collisions, items which I will have no time to cover but which
I want to point out to the interested reader. The first is the study of the
interplay between a finite top decay width and the spectrum of soft gluon
radiation emitted by the top itself and its decay products \cite{khoze}. The
predicted correlations are very difficult to observe, but perhaps future large
statistics of detected top and better confidence with new experimental
techniques will open new exciting possibilities. The second subject is a recent
calculation \cite{greco}\ of the resummed NLO high \pt\ behaviour of the  $b$
cross section, performed by matching the NLO fragmentation function formalism
\cite{nason}\ with the NLO massless matrix elements. The most important outcome
of this study is the reduced sensitivity of the spectrum on the choice of
renormalization and factorization scales.

\section{Charmonium Production}
Charmonium production in hadronic collisions has been measured both by UA1
\cite{ua1_psi}\ and CDF \cite{cdf_psi}, and has been studied within QCD for
quite some time now \cite{onia,gms}. However, all of the results available for
production at large \pt\ are relative to leading order calculations. Therefore
the theory of quarkonium production is not as solid as the theory of open heavy
quark production, and unfortunately a complete NLO calculation presents
serious difficulties and lies years ahead of us.

At leading order, large \pt\ charmonium is mostly
produced via gluon fusion diagrams such as the one shown in
fig.~\ref{ffragpsi}.
Simple power counting in the propagators suggests that the production at large
\pt\ is strongly suppressed: $d \sigma / d \pt^2 \sim 1/\pt^8$. This
suppression can be understood physically as a sort of form factor of the
charmonium state, form factor which is probed at $Q^2$ values of the order of
\pt: it is hard to accelerate a \jpsi\ to a large \pt\ without
destroying it in the process!

It was recently realised by Braaten and Yuan \cite{braaten}\ that there is a
class of higher order processes which are
not suppressed at large \pt\ as the leading order ones. These
are contributions where a high \pt\ gluon or charm quark will fragment into
charmonium states (Fig.~\ref{ffragpsi}).  While the probability that such a
parton fragments into, say, a \jpsi\ is very small, this probability is a
number relatively constant with \pt\ and the \jpsi\ will carry away a large
fraction of the parton momentum. Therefore the \pt\ spectrum of charmonium
states produced according to this fragmentation mechanism is similar to that of
the initial hard parton, namely something with an approximate shape $d \sigma /
d \pt^2 \sim 1/\pt^4$.

Fragmentation functions for charmonium production from gluon and charm quarks
have recently been calculated and their scaling violation studied
\cite{braaten}. They can be convoluted with the inclusive \pt\ distributions of
gluons and charm to give inclusive \pt\ distributions of charmonium states
\cite{fleming}. Some pieces are still missing, such as the fragmentation
production of $\chi$ states. Work on these is in progress, and for the time
being a final prediction for the spectrum of \jpsi's is still unavailable.

A complete fragmentation calculation is available however for the \psitwos,
which does not receive contributions from $\chi$ decays.
\begin{figure}
\begin{minipage}[t]{8.5cm}
\vskip 6.5cm
\caption[]{\tenrm \baselineskip=12pt
Comparison between different contributions to \psitwos\ production.
\label{fpsi2_th}}
\end{minipage}
\hskip 0.5cm
\begin{minipage}[t]{8.5cm}
\vskip 6.5cm
\caption[]{\tenrm \baselineskip=12pt
Azimuthal correlations between $b$ pairs, as a function of different triggers.
\label{fdphi}}
\end{minipage}
\end{figure}
In
fig.~\ref{fpsi2_th}\ I show the \psitwos\ \pt\ distributions predicted at 1.8
TeV from the three processes: direct LO production, gluon fragmentation and
charm fragmentation.  The fragmentation channels become larger than the LO
contribution as soon as $\pt>7$ GeV. In spite of this, however, the sum of all
these terms is still small compared to the effect of B decays, and small
compared to the current discrepancy observed between CDF data \cite{cdf_psi}\
and the sum of B decays and direct production. A similar problem is
apparent from the preliminary studies of the B fraction in \jpsi\ data
\cite{cdf_btau}, indicating a large contribution from non-B decays,
incompatible with  current estimates of direct \jpsi\ production.

It is clear that more has to be learned about charmonium production
in hadronic collisions. For example, there is a class of processes in between
the LO and the fragmentation ones, 
which behaves like $1/\pt^6$ and which probably contributes significatly to the
\pt\ region below 10-15 GeV. These processes have not been calculated yet. New
data and analyses dedicated to separate all possible sources of charmonium, for
example by separating samples on the basis of the \jpsi\ isolation and impact
parameter, will be fundamental to help theorists disentangle  what today seems
a rather embarassing issue for QCD.

\section{Heavy Quark Jets}
When discussing heavy quark production in hadronic collisions, is has become
customary to refer to properties of the heavy quark themselves rather than to
the properties of the jets in which they are embedded. For
this reason, available calculations are always formulated in terms of  the \pt\
spectrum of bottom mesons and not of their jets. Nevertheless, several
measures of current interest, say tagging of b-jets produced in association
with $W$ bosons, are formulated in terms  b-jet energies. What makes this
distinction interesting is the fact that different mechanisms will dominate
production of B mesons at a given \et\ and production of b-flavoured jets  of
the same \et.

With a language which is more pictorial than formal, we usually identify three
processes for heavy quark production: direct production, induced by $gg\to
Q\bar Q$ scattering, gluon splitting, given by a standard QCD $gg\to gg$
collision followed by the branching of a final state gluon into a  $Q\bar Q$
pair, and flavour excitation. The frequently asked question ``which is the
dominant process for $b$ production?'' admits several answers, depending on the
details of the events we are considering.

For example, say we are interested in the azimuthal correlations, \dphi. If we
consider a sample of events triggered by a 20 GeV $b$, we expect that phase
space will suppress in part configurations at \dphi=0 relative to $\dphi=\pi$.
In fact  configurations with \dphi=0 require the presence of a stiff gluon jet
recoiling against the $b\bar b$ pair, in order to compensate not only the
momentum of the triggered $b$, but also the momentum of the  $\bar b$ flying
parallel to the $b$.  We expect the overall \dphi\ distribution to be peaked at
$\dphi=\pi$.

If we trigger instead on a 20 GeV b-jet, namely a 20 GeV jet which contains a b
quark, regardless of its momentum, configurations with \dphi=0 and $\dphi=\pi$
have the same phase space, as in both cases it is sufficient the presence of
just two 20 GeV jets: in one case the first jet is made of the  $b\bar b$ pair
and the second jet is a gluon; in the other both jets are b jets from the
direct production channel. The dynamics of the collinear gluon splitting
enhances the \dphi=0 region, compensating the presence of an additional power
of \as, and one expects the resulting  distribution to have a double peak
structure, with peaks at \dphi=0 and $\pi$.

All of these considerations are qualitative, and will vary with energy and
other boundary conditions. Nevertheless one can observe their relevance by
looking at fig.~\ref{fdphi}, where the two different
cases are displayed for 20 and 50 GeV jets. The calculation was performed using
the fully exclusive NLO calculation by MNR \cite{mnr}, which allows to define
b-jets using the standard cone algorithms applied to inclusive jets in hadronic
collisions. Here we define ``jet'' a set of one or two partons\footnote{At the
NLO partonic jets contain at most 2 partons, as there are no more than 3
partons in the final state.} with $\Delta R<0.7$ ($\Delta
R=\sqrt{\Delta\eta^2+\Delta\phi^2}$). More specifically a b-jet is a jet in
which at least one of the forming partons is a $b$, while a gluon jet is a jet
just made out of one gluon. At the NLO, the \pt\ of a gluon jet is by
construction equivalent to the \pt\ of the heavy quark pair.

Using the work of ref.\cite{mnr}, one can calculate the b-jet inclusive \et\
distribution, and then answer at the NLO in QCD  another important question,
namely ``what is the fraction of jets of a given \et\ containing a heavy
quark''.  To do this it is sufficient to compare with the available NLO
calculations of the jet inclusive \et\ distributions \cite{nlojet}.
\begin{figure}
\begin{minipage}[t]{8.5cm}
\vskip 6.5cm
\caption[]{\tenrm \baselineskip=12pt \label{fbincet}
Inclusive \pt\ distribution of $b$ quarks, $b$-jets and gluon jets in
$b$ events. }
\end{minipage}
\hskip 0.5cm
\begin{minipage}[t]{8.5cm}
\vskip 6.5cm
\caption[]{\tenrm \baselineskip=12pt \label{fbfrac}
Fraction of jets containing a $b$ as a function of jet \et\
(solid); ratio between the $b$ \pt\ and inclusive jet \et\ spectra.}
\end{minipage}
\end{figure}
Fig.~\ref{fbincet}\ plots three NLO quantities: the inclusive $b$  (and $\bar
b$) \pt\ distribution (i.e. the NDE spectrum), the gluon jet \pt\ distribution
and the b-jet inclusive \et\ distribution. We applied no $\eta$ cuts in these
plots.  Taking the ratio of the heavy quark jet \et\ distribution with the
inclusive jet \et\ distribution, we obtain the curves shown in
fig.~\ref{fbfrac}, for charm and bottom jets. As a denominator, we used a LO
calculation for the inclusive jet \et\ spectrum. For $\Delta R=0.7$ and
$\mu=\pt/2$ this was shown to lead to a result numerically equal to the NLO
result. The results are consistent with indipendent estimates of the heavy
quark content of gluon jets \cite{hvqinjets}, and are consistent with
calculations performed \cite{unal}\ using the shower Monte Carlo HERWIG
\cite{herwig}. The flattening of the fractions at large \et\ reflects the
diminuishing percentage of gluonic jets in the inclusive jet sample with
larger \et.

Since, as observed above, measurements based on the b \pt\ or b-jet \et\
emphasize differently the various production mechanisms, a direct measurement
of the b-jet \et\ distribution would provide a check of the NLO calculations
which is complementary to the one performed using the standard NDE spectrum and
the B meson \pt\ distribution.

\subsection{Heavy Quark Jets and Photons}
Associated production of photons and charm quarks has been suggested as a probe
to study the charm density inside the proton \cite{halzen}.
The dominant production process is in fact $g c\to \gamma c$.
The similar process $g b\to \gamma b$ would allow the study of the $b$ density.
Additional interest in this process arises because it is a potential background
to the observation of the neutral current decay of a 4-th generation
$I_3=-1/2$ quark, $b^{\prime}\to b\gamma$ \cite{hou}. Finally, these processes
provide a useful control sample for the study of heavy quark tagging.

These processes are theoretically interesting because the density
inside the proton of a heavy quark is in principle calculable perturbatively
\cite{tung}.
Neglecting higher order logarithmic corrections, which can be resummed using
the Altarelli Parisi evolution, the inclusive process $p\bar p \to \gamma Q +X$
can be calculated by evaluating the partonic process $gg\to Q \bar Q\gamma$ and
integrating over the phase space of the $\bar Q$. This process is dominated by
configurations where the quark being integrated over is produced at large
rapidity and small \pt. No divergence will appear, because of the heavy quark
mass. A consistent definition of the heavy quark density, including thresholds
effects, should then reproduce the result of this calculation. Comparison
against the experimental data is however important, to verify that no
additional non-perturbative effect is at work.

\begin{figure}
\begin{minipage}[t]{8.5cm}
\vskip 6.5cm
\caption[]{\tenrm \baselineskip=12pt \label{fgamma}
Photon \pt\ distribution in events with a central $b$-jet
in direct photon events,
showing the results of the LO structure function approach and of the NLO
calculation. }
\end{minipage}
\hskip 0.5cm
\begin{minipage}[t]{8.5cm}
\vskip 6.5cm
\caption[]{\tenrm \baselineskip=12pt \label{fgammatyp}
Jet-type composition in $\gamma$-jet events.}
\end{minipage}
\end{figure}
I show in fig.~\ref{fgamma}\ the photon \pt\
distribution in events with a central $b$, calculated using the structure
function approach ($bg\to b\gamma$, solid line, CTEQ PDF's \cite{cteq}) and the
exact matrix elements for the $gg+q\bar q\to\gamma Q\bar Q$ processes, where
higher order corrections to the initial state evolution have not been included.
The $q\bar q$ channel produces a heavy quark pair via gluon splitting, and
cannot be accounted for by the structure function calculation. As the plot
shows there is perfect agreement between the two approaches, at least in the
region where gluon splitting is small. This indicates that the effects of
initial state evolution for the b quark at these values of $x$ and $Q^2$ are
not important. Notice that this conclusion is important for the consistency of
the NLO evaluation of the $b$ cross sections, which is performed without
including flavour excitation diagrams \cite{nde}, already accounted for at this
order in perturbation theory by the NLO matrix elements.
Similar results are obtained in the case of charm production, where however
there is some larger sensitivity to the choice of the charm mass.

Fig.~\ref{fgammatyp}\ finally
contains the distributions of jets of various type produced with photons.
Because of the difference in charge and partonic density, associated production
of $b$ is suppressed by a factor equal to 8--10 w.r.t. charm.
As a curiosity, notice that because of the suppression of the light-quark
annihilation channel, it is more likely for a jet produced in association with
a
photon to be a charm jet rather than a gluon jet, at least for trasverse
momenta up to 30 GeV.

\subsection{Associated Production of $W$ and $b$-jets}
This is a hot subject in these days, considering its implications on the search
for the top quark. A detailed study can be found in ref.~\cite{mlmwbb}. I will
present here some additional remarks, mostly matured from the work of and
discussions with various CDF collaborators, in particular
G. Unal and B. Williams \cite{unal}.

Given the importance of top search, it is important to be able to perform
estimates of the backgrounds which are as much as possible independent of
absolute normalizations provided by a theoretical calculation. The most
straightforward approach would be therefore to simply assume that the fraction
of b-jets in the inclusive jet sample is the same as in the $W$+jets sample.
The first can be measured, and the rate as a function of jet \et\ and possibly
other jet features can be then applied to the $W$+jet sample (see Contreras
presentation at this Workshop). In this fashion any theoretical uncertainty on
the absolute rate of inclusive $W$+multi-jet would drop out, being replaced by
the observed cross sections.

A priori there are several reasons to expect this assumption to be too naive.
First of all the inclusive jet sample has a large contribution from direct
production of $b$ jets, contribution which is absent in the case of the $W$
sample, where all heavy quarks come from gluon splitting \cite{mlmwbb}.
Secondly, the average fraction of gluon-initiated final state jets is
different in the two processes. This difference will affect the rate of heavy
quarks produced via gluon splitting. Both these reasons point to a smaller
fraction of heavy quark jets in the $W$ sample than in the inclusive jet
sample, indicating that at worst the above background assumption was
conservative.
Finally, even gluon jets of the same \et\ in the two samples will have a
different probability to produce $Q\bar Q$ pairs, because of the intrinsically
larger $Q^2$ scale of the collision producing a massive $W$.

In spite of these
differences, we expect the calculation of the relative
fractions of heavy quark jets to be less affected by theoretical uncertainties
than the absolute rates. The predictions can be tested on the
inclusive sample of $W$+ one $b$-jet + X, which is not significantly
contaminated by top. Notice that for this calculation it is fundamental to keep
into account the effects of the $b$ mass, in order to avoid divergencies
associated with the integration over the second $b$ phase space \cite{mlmwbb}.

\begin{figure}
\begin{minipage}[t]{8.5cm}
\vskip 6.5cm
\caption[]{\tenrm \baselineskip=12pt \label{fwbb}
Fraction of $W$+multijet events containing a $b$-jet with \et$>$20 GeV,
as a function of the number of jets above a given threshold. }
\end{minipage}
\hskip 0.5cm
\begin{minipage}[t]{8.5cm}
\vskip 6.5cm
\caption[]{\tenrm \baselineskip=12pt \label{fnjet}
Same for
inclusive multi-jet events. All jets have $|\eta|<2$.}
\end{minipage}
\end{figure}
Without getting into the details of the calculation, performed using a mixture
of the matrix element results of ref.~\cite{mlmwbb}\ and HERWIG \cite{herwig},
I show an indicative result in fig.~\ref{fwbb}, which gives the fraction of
$W$+multijet events containing a $b$-jet above 20 GeV, as a function of the
multiplicity of jets above a given \et\ threshold. It appears that the
probability per jet of being a $b$-jet is of the order of 1\%, almost
independently of the jet multiplicity.
For comparison, I show in fig.~\ref{fnjet}\ the same distribution obtained for
generic multi-jet events.  As expected, the fractions here are slightly larger,
indicating that indeed the naive assumption that jets in the two samples have
the same $b$-quark content is conservative.
\\[0.5cm]
{\bf Acknowledgements}:
I wish to thank my collaborators on heavy quark physics,
namely U. Baur, E. Braaten, M. Doncheski, S. Fleming, S. Frixione, S. Keller,
P. Nason and G. Ridolfi. I also wish to thank my collaborators at CDF for the
constant stimulus provided by their terrific work.

\end{document}